\documentclass{article}
\usepackage{epsfig}
\usepackage{graphicx}
\begin{document}

\title{\bf
Non-classical photon pair generation in atomic vapours }
\author{S. N. Sandhya\\
Department of Physics,Indian Institute of Technology\\ 
         Kanpur-208016, INDIA\\
email: sns@iitk.ac.in}
\maketitle
\begin{abstract}
A scheme for the generation of non-classical pairs of photons in atomic vapours is proposed.
The scheme exploits the fact that the cross correlation of the emission of photons from the extreme transitions
 of a four-level cascade system shows anti-bunching which has not been reported
earlier and which is unlike the case of the three level cascade emission
which shows bunching. The Cauchy-Schwarz inequality which is  
the ratio of cross-correlation to the auto correlation function in this case is estimated to be $10^3-10^6$ for controllable time delay,  and is one to four orders of magnitude larger compared to
previous experiments. The choice of Doppler free geometry in addition to the 
fact that at three photon resonance the excitation/deexcitation
processes occur in a very narrow frequency band, ensures cleaner signals.
\end{abstract} 

\section{Introduction}
Non-classical sources of light include triggered single photon emitters \cite{brouri}
as well as  photon pair emitters \cite{chou} where the first photon heralds the arrival of the second one.
The recent elegant experiments on the generation of non-classical photon pairs have
successfully reported the temporal control of emission of an anti-stokes
photon 
by programmable time delay. Polyakov et al \cite{polyakov},  report a violation of the classical Cauchy-Schwarz inequality 
in the photon correlations,  and they further report that the ratio of cross-correlation to auto-correlation
$R^{max}=292\pm 57$. All these experiments have been done on cold atoms in
magneto-optical traps (MOT)
. Recently, this experiment has been repeated in atomic vapours at room
temperature \cite{jiang} and the results are comparable with the MOT results. While in the MOT 
set up the magnetic and quadrupole fields induces decoherence in the correlations, in the latter case the inefficient absorption/excitation due to Doppler
broadening is the limiting factor. 

In this paper I propose a scheme for the generation of
non-classical pair of photons wherein Doppler effects are minimised even at room temperatures,
and the degree of violation of the Cauchy-Schwarz inequality is very large $\sim 10^3-10^6$.
The key feature we exploit here hinges on the fact that the
cross correlation between the emission of the extreme transitions in a four-level cascade system shows anti-bunching, 
which is in contrast to the well known fact that cross-correlation of the emission  
from  three level cascade systems shows bunching \cite{loudon,clauser}.
In addition, one can obtain very strong correlations between the photon emissions
from the extreme transitions by controlling the driving fields. 
To achieve this I consider here a model which consists of the 4-level cascade system in Rubidium which is an extension of the well studied 3-level cascade
system in the context of 
the absorptive as well as the dispersive property in electromagnetically induced transparency (EIT) systems \cite{banacloche, fulton, eit}.
In addition to the 3-level 
cascade system consisting of $5s_{1/2}, 5p_{3/2}$ and $5d_{5/2}$ considered in
 the previous examples, an additional level, say, one of the other hyperfine levels
 of $5p_{3/2}$ may be included and a strong coupling introduced between different
hyperfine levels of $5p_{3/2}$ through 
a radio-frequency (rf) field (see Fig1). The role of this 'sandwich' coupling is to change 
the usual 'bunched' emission from the cascade system to antibunched emission, with the difference
that the cascade emission is interrupted by the rf field after the first emission. This control field helps in inducing strong correlations
between the uppermost and the lower most transitions. 
It further provides us with  a handle to control the time delay between the successive emissions of photons
from the extreme transitions.
The combined feature of the violation of two classical inequalities
namely (i) antibunching 
and (ii) 
Cauchy-Schwarz inequality in this scheme makes it very attractive. Further the problem of the absorption/emission
into other frequency modes close to resonance (due to Doppler broadening)
which leads to the weakening of the signals \cite{jiang} is circumvented in the present
scheme because of the intrinsic narrow absorption features due to
atomic coherence effects induced by three-photon resonance.
 
The plan of the paper is as follows: In section II the model under consideration is described in detail and the equation of motion for the atom+field system is set up.
In section III the relevant photon correlation functions are evaluated numerically and in
some special cases analytical expressions for the correlation functions
are presented, which are evaluated perturbatively. Further, these results 
are compared with the exact numerical results for the correlation functions.

\section{The model}
 The system  considered here consists of an ensemble of four-level ladder
atoms (Fig1). This could correspond, eg., to $5s_{1/2}$, two hyperfine levels of $5p_{3/2}$ and $5d_{5/2}$ of Rb \cite{rubidium}. I would like to recall here that
this was exactly the scheme used by Banacloche et al \cite{banacloche}
and Fulton et al \cite{fulton} for studying 
EIT.
Labeling the energy levels $5s_{1/2}, 5p_{3/2}$ as $|1>$ and $|2>$ respectively,
level $|3>$ may be identified as one of the hyperfine levels of $5p_{3/2}$ (F=0,1,2,3 of $^{87}$Rb and F=1,2,3,4 of $^{85}$Rb ),
with level separations typically of the order of $10^2 MHz$. Level $|4>$ may be chosen to be
$5d_{5/2}$. 
 The only dipole allowed transitions are $|1>\leftrightarrow|2>$, 
$|2> \leftrightarrow |3>$ and $|3>\leftrightarrow |4>$ in the limit of the hyperfine splittings
being larger compared to the Rabi frequency \cite{shore}. Comparing this scheme with the
3-level cascade system of references \cite{banacloche,fulton,eit},
levels $|1>, |2> ,|3> $ of these setups correspond respectively to levels$ |1>,|2>$ and $
|4>$ of the present model. In addition we have introduced another hyperfine level
which we have labeled as $|3>$. 
\begin{figure}
\centering
\includegraphics[width=6cm]{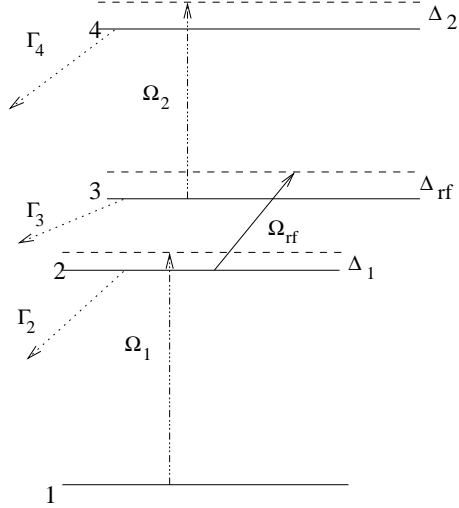}
\caption{ Four-level system interacting with three driving
fields of strengths $\Omega_1, \Omega_{rf},\Omega_3$ and their respective
detunings $\Delta_1,\Delta_{rf}$ and $\Delta_2$.
$\Gamma_2,\Gamma_3,\Gamma_4$ are the decay constants of the corresponding levels.
}
\end{figure}

For this system, the two transitions $|1>\rightarrow 2>$ and $|3>\rightarrow 4>$  respectively may be chosen to correspond to wavelengths 780nm and 775.8nm, 
as in the schemes used in references \cite{banacloche, fulton, eit}. Apart from these two driving fields in the optical
region, we consider yet another coupling between the hyperfine levels ( belonging to $5p_{3/2}$) $|2>$ and $|3>$ through
a rf field of strength $\Omega_{rf}$. Such couplings between the hyperfine levels
have been considered in the study of absorption in 4-level systems \cite{ suter}. 
 The strengths of the optical fields are denoted by $\Omega_1$ and $\Omega_3$, and the respective detunings by $\Delta_1$ and $\Delta_3$. 
The corresponding decay parameters are \cite{rubidium} $\Gamma_{2,3}=6 MHz$, $\Gamma_4\approx 1 MHz (0.97 MHz)$.
To make the three photon interaction Doppler free we choose the two beams 
in the optical region, as already mentioned, to be counter-propagating.
They  couple respectively transitions $|1> \leftrightarrow |2>$ and $|3> \leftrightarrow |4>$.
The optical frequencies, being nearly equal,
will give null contribution to the linear Doppler shift.
While, the linear Doppler shift of the rf field (in the atomic rest frame) at this frequency is very small and hence negligible. 
This choice for minimizing the Doppler effect is preferable
compared to the phase-matched geometry used in four-wave mixing as this offers a larger interaction 
volume compared to the latter. 

The master equation describing the interaction of the four-level atomic system with three driving fields in the rotating wave approximation is given by
\begin{eqnarray}
\frac {\partial {\rho}_{12}}{\partial t} =& (-i \Delta_1 -\Gamma_{2}/2)
{\rho_{12}} -i \Omega_1 ({\rho_{22}}-{\rho_{11}}) + i \Omega_{rf}
{\rho_{13}} \nonumber \\ 
\frac {\partial {\rho}_{23}}{\partial t} =& (-i \Delta_2 -(\Gamma_{2}+\Gamma_3)/2)
{\rho_{23}} -i \Omega_1 {\rho_{13}}- \nonumber \\
&i \Omega_{rf} ({\rho_{33}}-{\rho_{22}}) +
 i \Omega_3 {\rho_{24}} \nonumber \\
\frac {\partial {\rho}_{34}}{\partial t} =& (-i \Delta_3 -(\Gamma_{3}+\Gamma_4)/2)
{\rho_{34}} -i \Omega_{rf} {\rho_{24}} \nonumber \\
&-i \Omega_3 ({\rho_{44}}-{\rho_{33}}) 
 \nonumber \\
\frac {\partial {\rho}_{13}}{\partial t} =& (-i( \Delta_1+\Delta_2) -\Gamma_{3}/2)
{\rho_{13}} -i \Omega_1 {\rho_{23}} \nonumber \\
&+i \Omega_{rf} {\rho_{12}}+
i \Omega_3 {\rho_{14}}) \nonumber\\ 
\frac {\partial {\rho}_{14}}{\partial t} =& (-i( \Delta_1+\Delta_2+\Delta_3) -\Gamma_{4}/2)
{\rho_{14}} -i \Omega_1 {\rho_{24}} \nonumber \\
&+i \Omega_3 {\rho_{13}}
 \nonumber \\
\frac {\partial {\rho}_{24}}{\partial t} =& (-i( \Delta_2+\Delta_3) -(\Gamma_{2}+\Gamma_4)/2)
{\rho_{24}} -i \Omega_1 {\rho_{14}}\nonumber \\
&-i \Omega_{rf} {\rho_{34}}+
i \Omega_3 {\rho_{23}}) \nonumber \\
\frac {\partial {\rho}_{22}}{\partial t} =&  -\Gamma_{2} {\rho_{22}}+
i \Omega_1 ({\rho_{21}}-{\rho_{12}}) \nonumber \\
&+ i \Omega_{rf}
({\rho_{23}}-{\rho_{32}})+ \gamma_{23} {\rho_{33}} + \gamma_{24} 
{\rho_{44}} \nonumber \\
\frac {\partial {\rho}_{33}}{\partial t} =&  -\Gamma_{3} {\rho_{33}}+
i \Omega_3 ({\rho_{34}}-{\rho_{43}})\nonumber \\
& - i \Omega_{rf}
({\rho_{23}}-{\rho_{32}})+ \gamma_{34} {\rho_{44}}  
 \nonumber \\
\frac {\partial {\rho}_{44}}{\partial t} =&  -\Gamma_{4} {\rho_{44}}-
i \Omega_3 ({\rho_{34}}-{\rho_{43}}) \nonumber 
\label{spc}
\end{eqnarray}
where $\gamma_{ik}$ are the transition rates from the level $|k>$ to $|i>$,
$\Delta_i=\omega_{i,i+1}-\omega_i$ are the laser detunings
(${ \rho_{ij}} = { \rho_{ji}}^{*}$ and ${\rm Tr}{ \rho}=1).$ Throughout this paper we assume $\gamma_{i i+1}=1$ and the rest to be zero.
I obtain analytical solutions by perturbatively solving these equations
in the Laplace space in the special
cases where the rf field strength is either i) large, or ii) small compared to the
other two driving fields.

\subsubsection*{Case (i) Large $\Omega_{rf}$}
In this regime we assume that $\Omega_{rf} > \Omega_1, \Omega_3$. Treating
 the fields $\Omega_i,i=1,3$ perturbatively, and labeling
$ \psi_1={ \rho_{12}},\psi_2={ \rho_{23}},\psi_3={\rho_{34}},
\psi_4={ \rho_{13}},\psi_5={ \rho_{14}},\psi_6={ \rho_{24}},
\psi_7={ \rho_{22}},\psi_8={ \rho_{33}},\psi_9={ \rho_{44}}$ and $ \bar \psi_i(s)$ to be the Laplace transform of $\psi_i(t)$, 
the equation of motion in the Laplace
space assumes the simple form 
\begin{eqnarray}
(s+\bar \Gamma_2) \bar \psi_7^{(2)}(s)=&\psi_7(0)- 2 \Omega_{rf} Im \bar \psi_2^{(2)}(s)+\nonumber \\
&2 \Omega_1 Im \bar \psi_1^{(1)}(s)+ \bar \psi_8^{(2)}(s) \nonumber \\
(s+\bar \Gamma_2+\bar \Gamma_3) \bar \psi_2^{(2)}(s)=&- i \Omega_{rf}(\bar \psi_8^{(2)}(s)-\bar \psi_7^{(2)}(s)) -\nonumber \\
& i \Omega_1 \bar \psi_4^{(1)}(s)+i \Omega_3 \bar \psi_6^{(1)}(s) \nonumber \\ 
(s+\bar \Gamma_3) \bar \psi_8^{(2)}(s)=& \psi_8(0)+ 2 \Omega_{rf} Im \bar \psi_2^{(2)}(s)-\nonumber \\
&2 \Omega_3 \bar \psi_3^{(1)}(s)+\bar \psi_9^{(2)}(s)\nonumber \\
(s+\bar \Gamma_4) \bar \psi_9^{(2)}(s)=&\psi_9(0)+ 2 \Omega_3 Im \bar \psi_3^{(1)}(s)
\end{eqnarray}
where $\Omega_{rf}$ is treated upto all orders and 
the fields $\Omega_{1,3}$ are treated perturbatively. For simplicity I have assumed $\Delta_{rf}=\Delta_i=0, i=1,3$ and replaced $\Gamma_i/2$ by $\bar \Gamma_i$ 
everywhere.
Further, the first order solutions for $\psi_1^{(1)}, \psi_3^{(1)}, \psi_4^{(1)} {\rm and} \psi_6^{(1)}$
are obtained by solving
\begin{eqnarray} 
(s+\bar \Gamma_2) \bar \psi_1^{(1)}(s)=
& i \Omega_{rf} \bar \psi_4^{(1)}(s)+ 
i \Omega_1 (2 \bar \psi_7^{(0)}(s)+\nonumber \\
&\bar \psi_8^{(0)}(s)+\bar \psi_9^{(0)}(s)-1)\nonumber \\
(s+\bar \Gamma_3) \bar \psi_4^{(1)}(s)=&i \Omega_{rf} \bar \psi_1^{(1)}(s)\nonumber \\
(s+\bar \Gamma_4) \bar \psi_6^{(1)}(s)= &-i \Omega_{rf} \bar \psi_3^{(1)}(s)\nonumber \\
(s+\bar \Gamma_3+\bar \Gamma_4)\bar \psi_3^{(1)}(s)= 
&i \Omega_3 (\bar \psi_9^{(0)}(s)-\bar \psi_8^{(0)}(s))+\nonumber \\
&-i \Omega_{rf} \bar \psi_6^{(1)}(s) 
\end{eqnarray}
The solution for $ \rho_{ii}(t)$ depends on the choice of the initial condition which will be discussed in section III.
\subsubsection*{
Case (ii) Small $\Omega_{rf}$.}
I next consider the case when the rf field is weak. Considering the contribution of $\Omega_{1,3}$ upto all orders and treating $\Omega_{rf}$ perturbatively,the second order equations are given by
\begin{eqnarray}
(s+\bar \Gamma_2)\bar \psi_7^{(2)}=&2 \Omega_1 Im \bar \psi_1^{(2)}-2\Omega_{rf} Im \bar \psi_2^{(1)}+\nonumber \\
&\bar \psi_8^{(2)}+\psi_7(0)\nonumber \\
(s+\bar \Gamma_3) \bar \psi_8^{(2)}=& -2 \Omega_3 Im \bar \psi_3{(2)}+2 \Omega_{rf} Im \bar \psi_2^{(1)}+\nonumber \\
&\bar \psi_9^{(2)} + \psi_8(0)\nonumber \\
(s+\bar \Gamma_4) \bar \psi_9^{(2)}=& 2 \Omega_3 Im \bar \psi_3^{(2)}+ \psi_9(0)\nonumber \\
(s+\bar \Gamma_3+\bar \Gamma_4) \bar \psi_3^{(2)}=&-i \Omega_{rf} \bar \psi_6^{(1)}- i \Omega_3 (\bar \psi_9^{(2)}- \bar \psi_8^{(2)}) \nonumber \\
(s+\bar \Gamma_2)\bar \psi_1^{(2)}=& -i\Omega_1(2 \bar \psi_7^{(2)}+\bar \psi_8^{(2)}+\bar \psi_9^{(2)}-1)+\nonumber \\
& i\Omega_{rf} \bar \psi_4^{(1)}
\end{eqnarray}
where the first order $\bar \psi_i$ are obtained by solving 
\begin{eqnarray}
(s+\bar \Gamma_3+\bar \Gamma_2)\bar \psi_2^{(1)}=&-i \Omega_1 \bar \psi_4^{(1)}+ i \Omega_3 \bar \psi_6^{(1)}\nonumber \\
& -i \Omega_{rf} (\psi_8(0)-\psi_7(0)) \nonumber \\
(s+\bar \Gamma_3)\bar \psi_4^{(1)}=& -i \Omega_1 \bar \psi_2^{(1)} + i \Omega_3 \bar \psi_5^{(1)}\nonumber \\
(s+\bar \Gamma_2+\bar \Gamma_4)\bar \psi_6^{(1)}=& -i \Omega_1 \bar \psi_5^{(1)} + i \Omega_3 \bar \psi_2^{(1)}\nonumber \\
(s+\bar \Gamma_4)\bar \psi_5^{(1)}=& -i \Omega_1 \bar \psi_6^{(1)} + i \Omega_3 \bar \psi_4^{(1)}
\end{eqnarray}
The solutions of this equation are discussed in the next section.

\section{Correlation Function}
Our interest is in determining
the second order correlation function $G_{ij}^{(2)}({\bf r}_1,t_1,{\bf r}_2,t_2)$
which gives the correlation between the fluorescence signals $I_i$ and $I_j$ at times
$t_1=t-r_1/c$ and $ t_2=t-r_2/c$, and ${\bf r}_1, {\bf r}_2$ are the positions
of the detectors. Here i,j=1,3 corresponds to the emission of the g.s. and the
upper excited state respectively. The correlation function 
is determined by evaluating the expectation values of the product of two-time operators
$<E_i^-({\bf r}_1, t_1)E_j^-({\bf r}_2,t_2)E_j^+({\bf r}_2,t_2)E_i^+({\bf r}_1,t_1)>$ for i=1 and 3 (since we are interested only in the uppermost
and the g.s. transitions) where $E^+$ and $E^-$ are the positive and negative frequency
components of the field emitted. 

The normalized temporal intensity correlation function $g^{(2)}_{ij}(t_1,t_2)$ can be written in terms
of the atomic operators in the far-zone approximation \cite{gsa, charmichael} to be
\begin{equation}
g_{ij}^{(2)}(t_1,t_2)=\frac{<\sigma_i^+(t_1)\sigma_j^+(t_2)\sigma_j^-(t_2)\sigma_i^-(t_1)>}{<\sigma_i^+ \sigma_i^->_{ss}<\sigma_j^+ \sigma_j^->_{ss}}
\end{equation}
for unit detector efficiency, where $<\sigma_i^+ \sigma_i^->_{ss}$ corresponds to the steady state signal
of the ith mode, $\sigma_i^{+}=|i+1><i|$ and $ \sigma_i^{-}=|i><i+1|$ being the atomic operators. 
The two-time expectation values are evaluated using the well known Onsager-Lax Quantum
regression theorem \cite{lax} which states that if for a complete set of system operators ${\hat{\cal A}}_\mu , \mu = 1, 2...n$
the one time expectation value satisfies 
\begin{equation}
<{ \hat{ \cal A}}_\mu(t)>= \sum_{\mu} {\cal M}_{\mu \nu}(t,t')<{ \hat {\cal A}}_{\nu}(t')>, t'< t
\end{equation}
then the two-time expectation value takes the form
\begin{eqnarray}
<{ \hat {\cal O}}_{\alpha}(t') {\hat {\cal A}_{\mu}}(t){\hat {\cal O}}_{\beta}(t')>=&
\sum_{\nu}{\cal M }_{\mu \nu}(t,t')  \nonumber \\
&
<{\hat{ \cal O}}_{\alpha}(t'){\hat{ \cal A}_{\nu}}(t'){\hat {\cal O}}_{\beta}(t')>,\nonumber \\
& t'<t
\end{eqnarray}
for any two system operators ${\hat{\cal O}}_{\alpha}$ and ${\hat{\cal O}}_{\beta}$. Essentially the
correlation functions (two-time averages) satisfy the same equation of motion as that of the expectation values of one-time averages
\cite{charmichael} . The c-number coefficients ${\cal M}_{\mu \nu}$
are derived from the solution of the Heisenberg equation. Using the operator
algebra $\sigma_{ij}\sigma_{kl}=\delta_{jk}\sigma_{il}$ and Eqn.(5), 
the two-time averages
for the four-level cascade system may be expressed as
\cite{huang, jha, charmichael} 
\begin{eqnarray}
g^{(2)}_{11}(\tau)=& <\rho_{22}(\tau)>_{\rho(0)=|1><1|} / \rho_{22}^{ss}\nonumber\\
g^{(2)}_{33}(\tau)=& <\rho_{44}(\tau)>_{\rho(0)=|3><3|}/\rho_{44}^{ss}\nonumber \\
g^{(2)}_{31}(\tau)=& <\rho_{22}(\tau)>_{\rho(0)=|3><3|}/\rho_{22}^{ss}
\end{eqnarray}
\begin{figure}
\begin{center}
\includegraphics[width=8cm]{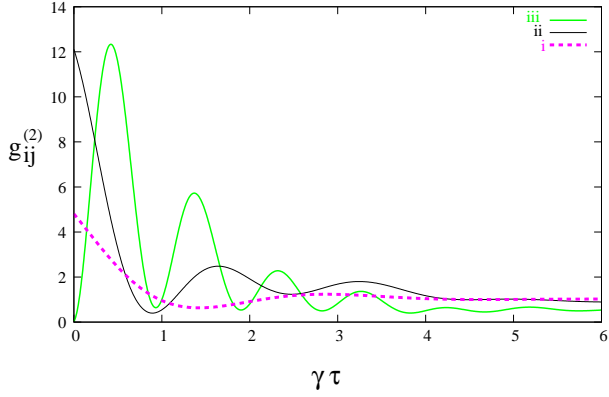}
\end{center}
\caption{(Color online) The cross-correlation functions a) $g^{(2)}_{31}(\tau)$, b) $ g^{(2)}_{32}(\tau)$,
and c) $ g^{(2)}_{21}(\tau)$ 
function of $\gamma \tau$
with $\Omega_1=\Omega_{3}=4{\gamma}$, and  
$\Omega_{rf}=20{\gamma}$.}
\end{figure}
\begin{figure}
\begin{center}
\includegraphics[width=8cm]{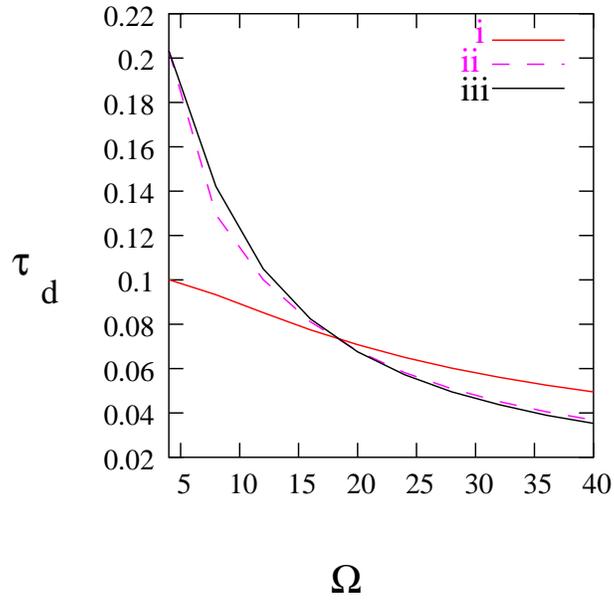}
\end{center}
\caption{(Color online) The time delay between the emission
of the photons from the extreme transitions as a function
of the driving field strengths
i) $\Omega_1$ with $\Omega_2=12\gamma$ and $\Omega_3=4\gamma$,
ii) $\Omega_2$ with $\Omega_1=\Omega_3=4\gamma$,
iii) $\Omega_3$ with $\Omega_1=\Omega_2=4\gamma$,
}
\end{figure}
\begin{figure*}
\begin{center}
\includegraphics[totalheight=2.2in]{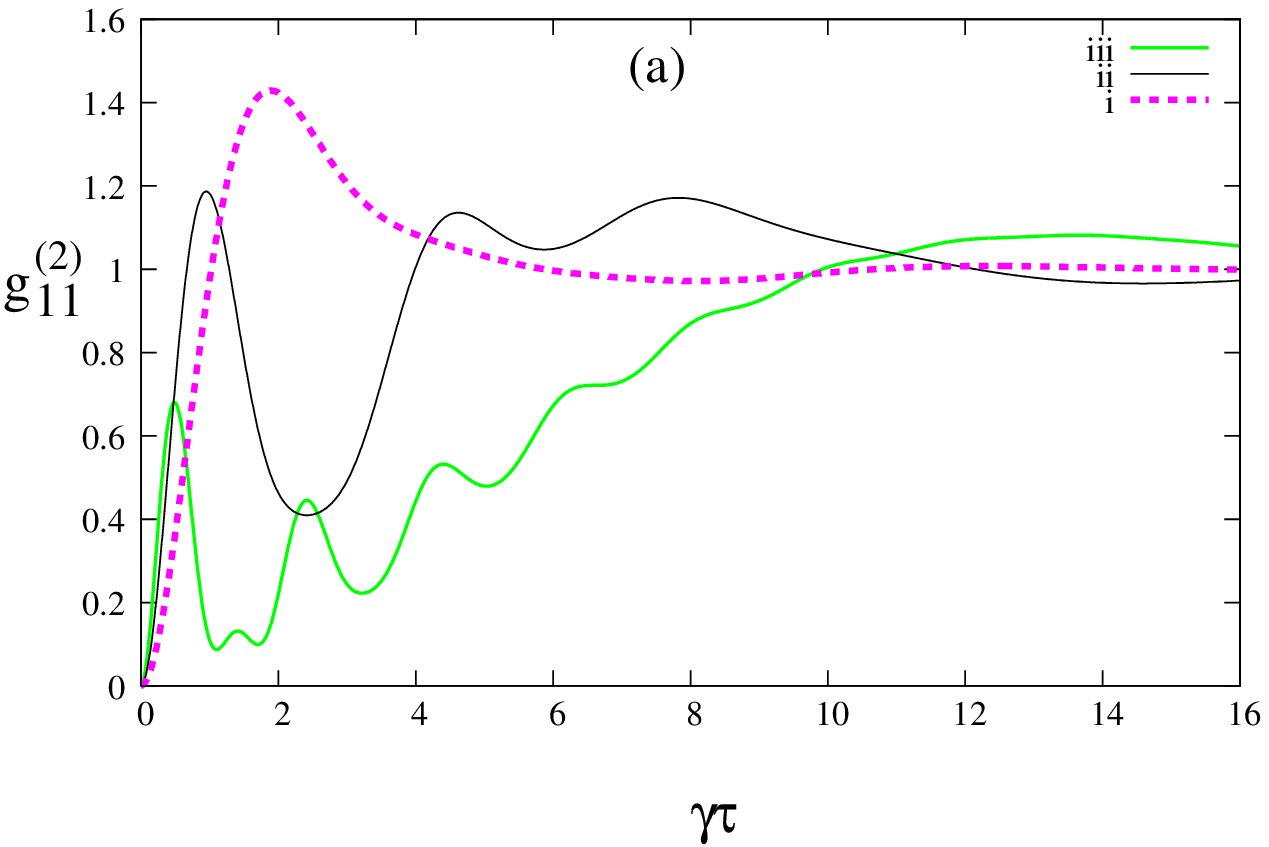}
\includegraphics[totalheight=2.2in]{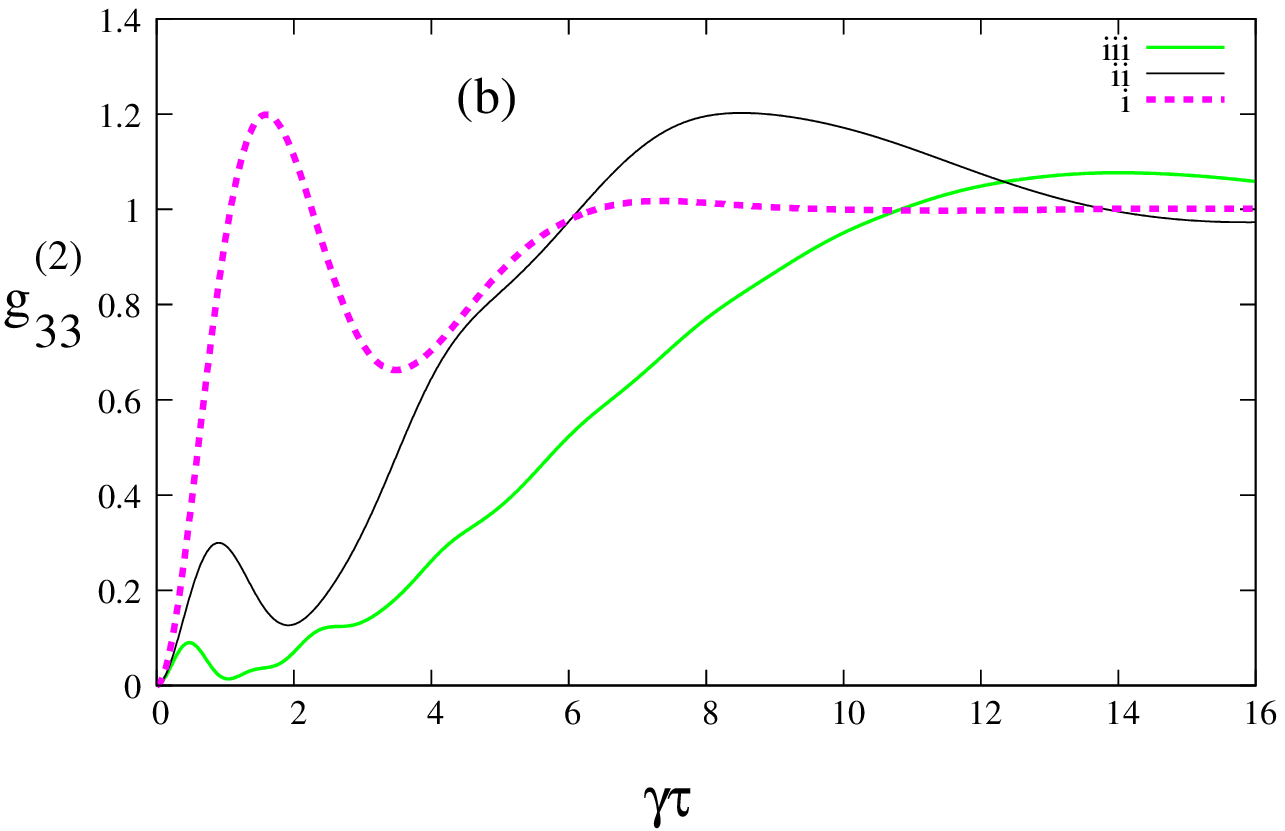}\\
\includegraphics[totalheight=2.2in]{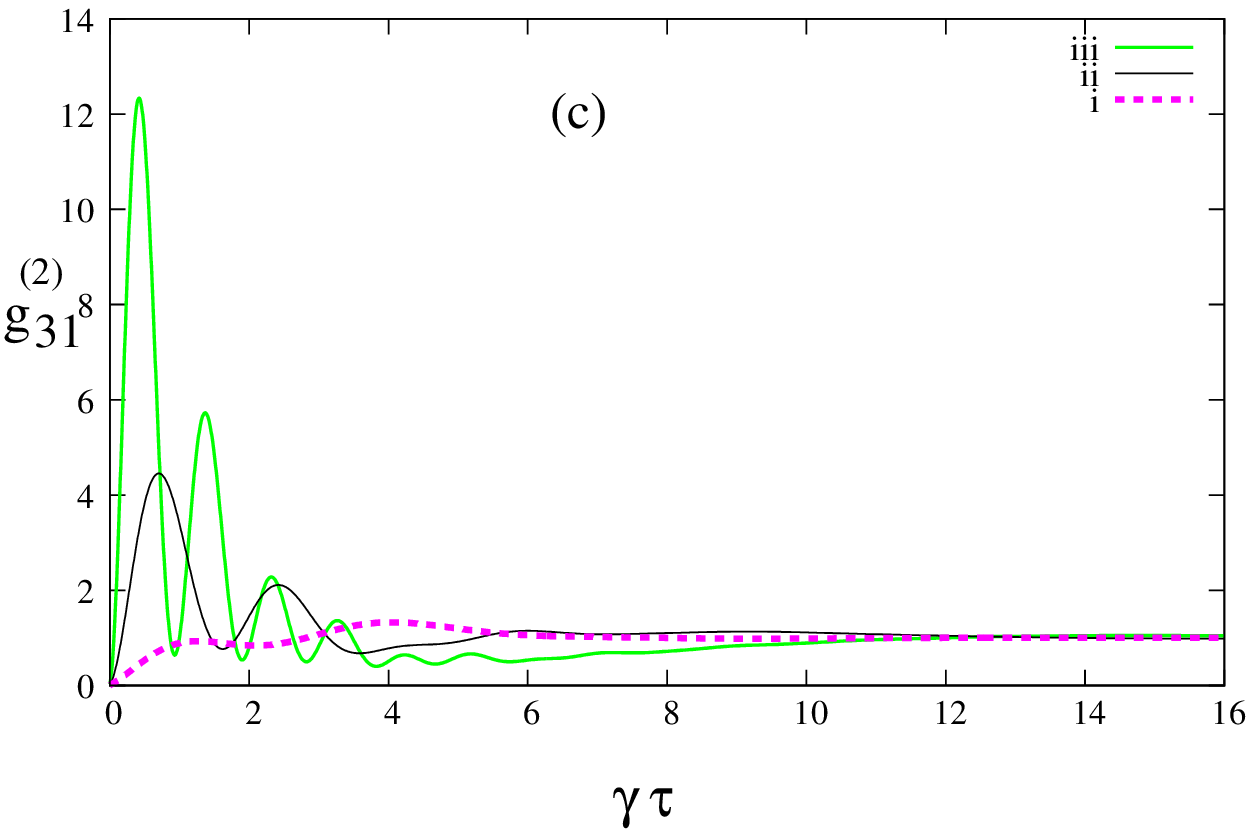}
\includegraphics[totalheight=2.2in]{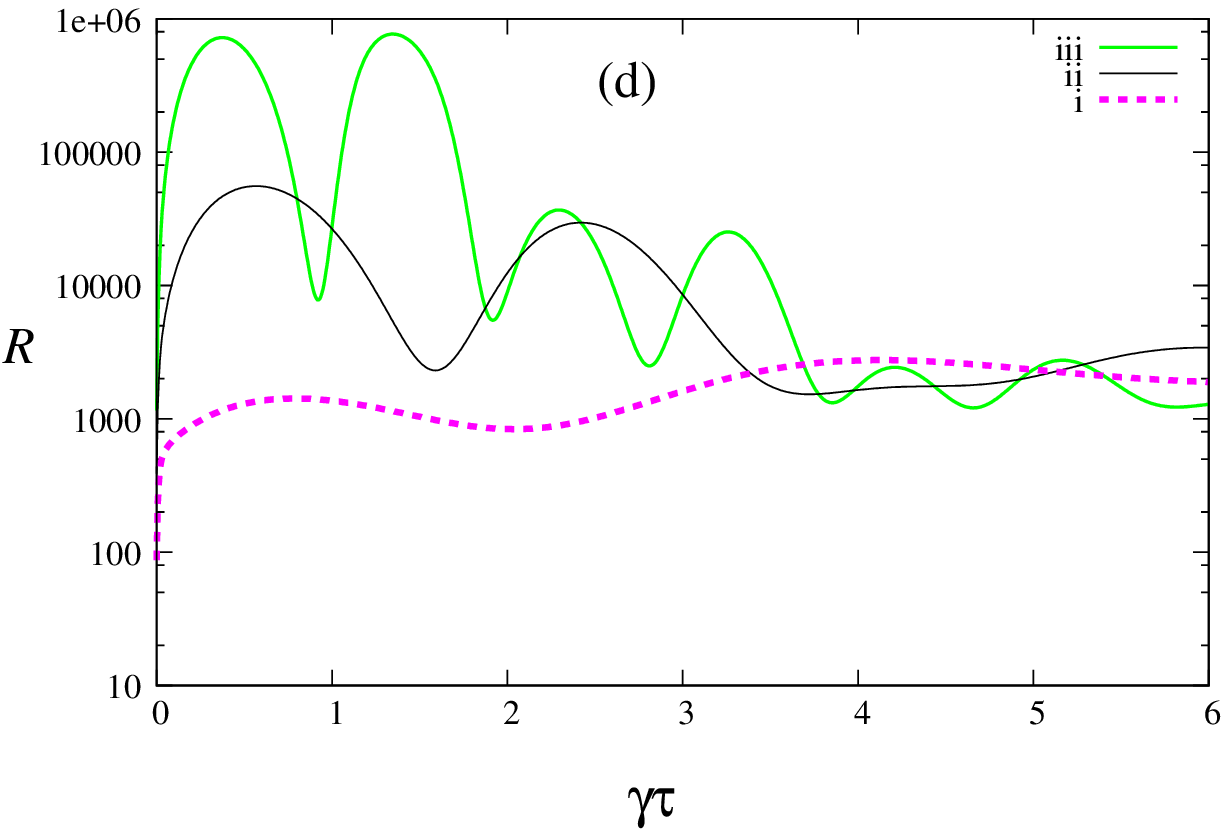}\\
\end{center}
\caption{(Color online) The correlation functions a) $g^{(2)}_{11}(\tau)$, b) $ g^{(2)}_{33}(\tau)$,
and c) $ g^{(2)}_{31}(\tau)$ and d) the ratio $R$ as a function of $\gamma \tau$
with $\Omega_1=\Omega_{3}=4{\gamma}$, and (i) $\Omega_{rf}=4{\gamma}$ ii) $\Omega_{rf}=10{\gamma} $ 
and iii) $\Omega_{rf}=20{\gamma}$.}
\end{figure*}
where $\tau=t_2-t_1$. Here $\rho_{ii}^{ss}$ denote the steady state
values. The functions $g^{(2)}_{11}(\tau)$  and $g^{(2)}_{33}(\tau)$ are the
auto-correlation functions of the g.s. emission  ($|1> \rightarrow |2>$) and the upper excited state emission ($ |4>\rightarrow|3>$) respectively. The cross-correlation function $g^{(2)}_{31}(\tau)$ gives the conditional probability 
for the emission of a photon from the $|1>\rightarrow |2>$ transition at a time $\tau$ given
an emission of a photon from the $|3>\rightarrow |4>$ transition at time $t=0$.
Here we have used the notation of reference \cite{charmichael}:
 $<\rho_{ii}(\tau)>_{\rho(0)=|j><j|}$ denotes the population $\rho_{ii}$ at a time
$\tau$ given the initial condition that the atom is prepared such that $\rho(0)=|j><j|$.
The time dependent 
functions $\rho_{ij}(\tau)$ in the general case is determined numerically
 while in some special cases I provide the
 analytical solutions. I first consider the case when $\Omega_{rf}$ is large. Solving the eq.(1) and (2)  in the Laplace space wherein both $\Omega_1$ and $\Omega_3$
 are treated perturbatively while $\Omega_{rf}$ is considered upto all orders
,  
the correlation functions are given by
\begin{eqnarray}
g_{11}^{(2)}(\tau)
&= 1+\sum_{i=1}^{5} {\bar a_i} e^{-\alpha_i \tau}\nonumber \\
g_{31}^{(2)}(\tau)
&=
 1+ {e^{-\bar \Gamma_4 \tau}}(\bar b_0+\bar b_1 e^{-\alpha_1 \tau}+
{\bar b_2} e^{-\alpha_2 \tau})+ \nonumber \\
&\sum_{i=3}^{5}{\bar b_i} e^{-\alpha_i \tau}
 \nonumber \\
g_{33}^{(2)}(\tau)
&= 
1+ e^{-\bar \Gamma_4 \tau}(\bar c_0+\bar c_1 e^{-\alpha_1 \tau}+
{\bar c_2} e^{-\alpha_2 \tau}) \nonumber \\
\end{eqnarray}

where $\bar a_i=a_i/a_N $
 $\bar b_i=b_i/b_N {\rm and} \bar c_i=c_i/c_N$ 
 is the normalisation factors $a_N, b_N, c_N$ . The coefficients are given by
\begin{eqnarray}
a_N=& 2\Omega_1^2 \Omega_{rf}^2/\Pi_{i=1}^5 (\alpha_i)\nonumber \\ 
 b_N=&2 \Omega_{rf}^2\Omega_3^2/\bar \Gamma_4 (\Pi_{i=1}^5 (\alpha_i))\nonumber \\
 c_N=& \alpha_0/\bar \Gamma_4 \alpha_1 \alpha_2\nonumber \\
c_0=& - \alpha_0/\bar \Gamma_4 (\bar \Gamma_4-\alpha_1)(\bar \Gamma_4-\alpha_2) \nonumber \\
 c_1=&- \alpha_0 /\alpha_2 (\alpha_1-\alpha_2)(\bar \Gamma_4-\alpha_2) \nonumber \\
c_2=&\alpha_0/\alpha_1(\alpha_1-\alpha_2)(\bar \Gamma_4-\alpha_1) \nonumber \\
 \alpha_0=& 2 \Omega_3^2(\bar \Gamma_2+\bar \Gamma_4)\nonumber 
\end{eqnarray}
The roots in the exponent are given by 
the conjugate pairs 
\begin{equation}
\alpha_{1,2}=-\bar \Gamma_2-\bar \Gamma_3 \pm i \phi ; \phi= \sqrt{4\Omega_{rf}^2-(\bar \Gamma_2-\bar \Gamma_3)^2}\nonumber
\end{equation}
and the cubic roots
\begin{eqnarray}
\alpha_3=&-\frac{2}{3\alpha}(\bar \Gamma_2+\bar \Gamma_3)-4\Omega_{rf}^2 +\frac{\alpha}{3}\nonumber \\
\alpha_{4,5}= &-\frac{2}{3\alpha}(\bar \Gamma_2+\bar \Gamma_3)+12 (1\pm i \sqrt{3})\Omega_{rf}^2 -\frac{1\mp i\sqrt{3}}{\alpha} \nonumber 
\end{eqnarray}
where
\begin{eqnarray}
\alpha=&3^{\frac{1}{3}}(3 \Omega_{rf}^2(\bar \Gamma_2+\bar \Gamma_3)+
\nonumber \\
&\frac{1}{6}\sqrt{324(\bar \Gamma_2+\bar \Gamma_3)^2\Omega_{rf}^4+6912 \Omega_{rf}^6})^{\frac{1}{3}}\nonumber
\end{eqnarray}
Note that the coefficients $c_i$ satisfy the condition
\begin{eqnarray}
  \sum_{i=0}^2 c_i + c_N =& 0 \nonumber 
\end{eqnarray}
The coefficients $a_i, b_i, i=1,5 $ are functions of $\alpha_i$ and since the expressions are very lengthy we have not listed them here.
However, the solution for $ \rho_{ii}(\tau)$ in the Laplace space which are displayed in the
Appendix, yield an important identity amongst the coefficients 
\begin{eqnarray}
a_N + \sum_{i=0}^5 a_i =& 0 \nonumber \\
b_N + \sum_{i=0}^5 b_i =& 0 \nonumber
\end{eqnarray}
 While the conditions for the coefficients $a_i$ and $c_i$  reflect the antibunching nature of the auto-correlation function,  
the condition for $b_i $ implies the anti-bunching nature of the
cross correlation function $g_{31}^{(2)}(\tau)$.
This contrasting feature which is exhibited by a 4-level ladder system is
unlike the behaviour of the cross correlation functions $ g_{21}^{(2)}(\tau)$ in a 3-level cascade
system which is non-zero at $\tau=0$ . A comparison of the various
cross-correlation functions of a 4-level cascade system will be discussed in
detail shortly.

The expressions in Eqn.(9) for the correlation functions indicate that the decoherence of $g_{33}^{(2)}(\tau),g_{31}^{(2)}(\tau)$ 
is controlled by the decay constants $\bar \Gamma_2, \bar \Gamma_3$  and $ \bar \Gamma_4$ 
while the dephasing of the g.s. autocorrelation function is due to $\bar \Gamma_2$
and $\bar \Gamma_3$ only.
The imaginary part in the exponents indicate that
the
the strength of the correlation oscillates with 
$\Omega_{rf}$. 

In the limit of weak $\Omega_{rf}$ the solutions for the correlation function are given by
\begin{eqnarray}
g_{11}^{(2)}(\tau)
&= 1-e^{-\bar \Gamma_2\tau}( Cos(2 \Omega_1 \tau)-\nonumber \\
&d_1 Sin(2 \Omega_1 \tau))\nonumber \\
g_{31}^{(2)}(\tau)
&= 1+e^{-\bar \Gamma_2 \tau}(x_0+
x_1 Sin(\Omega_1 \tau)+x_2 Cos(\Omega_1 \tau)+\nonumber \\
& \sum_{i=3}^{9}x_i e^{-\bar \alpha_i \tau} \nonumber \\
g_{33}^{(2)}(\tau)
&= 1+ \sum_{i=0}^{6} y_i e^{-\bar \alpha_{i+3} \tau}
\end{eqnarray}
 where $d_1=\bar \Gamma_2 d_0/2\Omega_1$,
with $d_0=2\Omega_1^2/(4\Omega_1^2+\bar \Gamma_2^2)$. 
 The roots $\bar \alpha_i,i=3,5 $ are obtained by replacing $\bar \Gamma_2 \rightarrow \bar \Gamma_3,
 \bar \Gamma_3 \rightarrow \bar \Gamma_4 {\rm and} \Omega_{rf} \rightarrow \Omega_3$ in
$\alpha_i$.
 Further
\begin{eqnarray}
\bar \alpha_{6,7}=&-(\bar \Gamma_2+\bar \Gamma_3+\bar \Gamma_4) \mp \sqrt{\phi_3 +2 \phi_1 \phi_2}\nonumber \\
\bar \alpha_{8,9}=& -(\bar \Gamma_2+\bar \Gamma_3+\bar \Gamma_4) \mp i\sqrt{ \phi_3- 2 \phi_1 \phi_2}\nonumber
\end{eqnarray}
where
\begin{eqnarray}
 \phi_3=& (4 (\Omega_1^2+\Omega_3^2)-(\bar \Gamma_2^2+\bar \Gamma_3^2+\bar \Gamma_4^2-2 \bar \Gamma_3 \bar \Gamma_4)\nonumber \\
\phi_1=& \sqrt{(4 \Omega_1^2-\bar \Gamma_2^2)}\nonumber \\
\phi_2=&\sqrt{4 \Omega_3^2- (\bar \Gamma_3-\bar \Gamma_4)^2}\nonumber
\end{eqnarray}
In this regime, the g.s. autocorrelation function resembles the second order correlation
function of the usual two-level system. This is intuitively obvious since the
upper level coupling is weak. Hence the dephasing of this function depends on $\bar \Gamma_2$.
The strength of the correlation oscillates with $\Omega_1$ as shown by equation (10).
While the dephasing of $g_{33}^{(2)}(\tau)$ and $g_{31}^{(2)}(\tau)$ depends on $\bar \Gamma_i, i=2,4$.
The strength of the correlation now depends on $\Omega_1$ and $\Omega_3$.

It follows from the Eqn.(10) that $g_{11}^{(2)}(\tau)=0$ for $\tau=0$. 
 Again, it may be verified using the solutions for 
$<\rho_{44}>_{\rho(0)=|3><3|}$ and $<\rho_{22}>_{\rho(0)=|3><3|}$ in the Laplace space given in the Appendix,
that the auto-correlation function $g^{(2)}_{33}(\tau)$ and the
cross- correlation function $g_{31}^{(2)}(\tau)$ are zero at $\tau=0$ . 
 Since the expressions for $x_i$ and $y_i$ are too lengthy,
we 
list all the solutions in the Laplace space in the Appendix.
Thus, in either of the regimes, I would like
to emphasize that 
the function $g^{(2)}_{31}(\tau)=0$ at $\tau=0$. This is further
evident from the numerical results which are solved exactly and 
valid for all strengths of the coupling fields (Fig4c).
Note that the cross-correlation is very strong for $\gamma \tau <1$.
Whenever the emission of the two photons occurs for time delay smaller than the 
lifetime of the atom, it necessarily means that both the photons belong
to the same cascade emission. Hence the photons show very strong correlation
for times smaller than $\gamma^{-1}$. For larger time delay between the
emissions of the two photons decoherence weakens the correlation.
\subsubsection*{Comparison of various cross-correlation functions}
For easy comparison with the 3-level cascade system, we consider
the most general 4-level system and label the 
emission from the $|i+1>\rightarrow |i>$ as the $i^{th}$ mode.
The cross-correlation functions $g_{ij}^{(2)}(\tau), i\ne j$ which are proportional to
$<\rho_{j+1,j+1}(\tau)>_{\rho(0)=|i><i|}$ listed in the Appendix,
give the probability of emission of photon in the $j^{th}$ mode at
a time $\tau$
given the emission of a photon in the $i^{th}$ mode at t=0.
The function $g_{21}^{(2)}(\tau)$ corresponds to the usual cross-correlation function of the 3-level cascade system and is nonzero at $\tau=0$. Again
$g_{32}^{(2)}(\tau)$ is also nonzero at $\tau=0$. This is easily understood
by looking at the solutions in the Laplace space (see Appendix).
Both $g_{32}^{(2)}(\tau)$ and $g_{21}^{(2)}(\tau)$ which are respectively proportional to
$<\rho_{33}(\tau)>_{\rho(0)=|3><3|}$ and $<\rho_{22}(\tau)>_{\rho(0)=|2><2|}$ have contribution from terms like $s^2/(s+\alpha_3)(
s+\alpha_4)(s+\alpha_5)$ in the Laplace space the inverse transform of which has the form
$\sum_{i}A_i e^{-\alpha_{i} \tau}$ where $A_i= \alpha_i^2/(
\alpha_i-\alpha_j)(\alpha_i-\alpha_{k}), i\ne j\ne k , i,j,k=3,5$. 
Clearly, at $\tau=0$, the $\sum_{i=3}^5 A_i \ne 0$ . 
I would like to recall here similar results due to Loudon \cite{loudon}
for a 3-level cascade system, where $g_{21}^{(2)}(\tau)$ is nonzero
at $\tau=0$ . 
On the other hand, the
distinguishing feature $g_{31}^{(2)}(\tau\rightarrow 0)= 0$ is due to the
absence of such terms.
Thus, while $g_{32}^{(2)}(\tau)$
and $g_{21}^{(2)}(\tau)$ both show bunching, $g_{31}^{(2)}(\tau)$ is
zero at $\tau=0$. The exact numerical solutions which are valid for all
strengths of the coupling fields, is illustrated in Fig2.

 To understand the 
underlying mechanism behind this feature consider the limit $\Omega_{rf}\rightarrow0$ and $\Omega_i, i = 1, 2 \rightarrow 0$. 
In this limit the correlation function $g_{21}^{(2)}(\tau)$ is proportional to $e^{-(\Gamma_2/2) \tau} $. i. e. the correlation for the two photon  emission pathway $|3>
\rightarrow |2> \rightarrow |1>$ depends on $e^{-(\Gamma_2/2) \tau}$. Likewise ,
the cross-correlation for the emission pathway $|4>\rightarrow |3> \rightarrow |2>$, $g_{32}^{(2)}(\tau)$ is proportional to $e^{-(\Gamma_3/2) \tau}$
and hence both these correlations are nonzero at $\tau=0$. On the other hand the $g_{31}^{(2)}(\tau)=
2/(\Gamma_3 -\Gamma_2)(e^{-(\Gamma_2/2) \tau} -e^{- (\Gamma_3/2) \tau})$. This implies that the cross correlation of the photons emitted by the extreme 
transitions gets contribution from both the two-photon pathways. 
At $\tau=0$ this function vanishes. In other words, there is an exact cancellation of emission due to the two pathways $|4>\rightarrow|3>\rightarrow|2>$
and $|3>\rightarrow |2>\rightarrow|1> $ at $\tau=0$. Thus the antibunching nature of $g_{31}^{(2)}(\tau)$ is intrinsic to the four-level cascade system and
is independent of the driving field strengths. Since this implies a finite
 time delay for the emission of the second photon and since the offset is 
provided by the strength of the driving fields (Fig4c),
in principle one has a handle on the control of this time delay $\tau_d$. 
Fig3 shows the variation of $\tau_d$ with $\Omega_i, i=1,3$ where $\tau_d$
is 
the time delay at which the cross-correlation peaks.
It is clear
from Fig3 that larger the field strengths, smaller the time delay
between the emissions.
As $\Omega_2$ or $\Omega_3$ is increased
the uppermost level gets populated much faster and hence the peak of the
cross-correlation function occurs at earlier $\tau_d$ ( This is reflected
in Fig4c (ii) and (iii) for $\Omega_{rf}= 10\gamma$ and $20\gamma$). While increasing
$\Omega_1$ increases the population in level $|2>$ till saturation. Since 
this does not affect the cross-correlation significantly, the variation
of $\tau_d$ with $\Omega_1$ is not as significant as with $\Omega_{2,3}$. 
The control will be 
better with the use of pulsed excitation with programmable time delay (of the order of a few $\gamma$'s) between
successive excitations \cite{chou, polyakov}. This provides an advantage over the other schemes
like the photon down conversion or the usual three level cascade emission
where the emission of both the photons is almost simultaneous (bunched) and hence
there is no way of delaying the emission of the second photon after the first photon is emitted.

I would like to recall here that 
antibunching is the violation of the classical
inequality: $g^{(2)}(0) \ge g^{(2)}(\tau)$. This inequality is satisfied by 
'classical fields', i. e., 
the  Glauber-Sudarshan
phase-space
function corresponding to the states of the field has positive distribution. 
Violation of these inequalities would imply that the 
distribution function corresponding to the state is not well behaved (not 
positive definite) 
and are non-classical.
Having demonstrated the anti-bunching nature of $g_{31}^{(2)}(\tau)$
in this section, 
I next discuss the violation of yet another classical inequality namely, 
the Cauchy-Schwarz inequality. 
\subsubsection*{Violation of the Cauchy-Schwarz Inequality}
 The other signature for the non-classical nature of the emitted fields
 is the violation of Cauchy-Schwarz inequality \cite{clauser}. The
degree of violation is determined 
by the ratio $R=(g^{(2)}_{31}(\tau))^2/g^{(2)}_{33}(0) g^{(2)}_{11}(\tau)$, $R \le 1$ for a classical source. One needs to optimize
the emissions in such a way that one achieves a large $R^{max}$.
In the present scheme, this condition essentially boils down to 
demanding sub-poissonian statistics in the $g^{(2)}_{ii}(\tau)$ for $i=1,3$
-which 
ensures that the photons are emitted at spaced out intervals - 
in addition to large values for $g^{(2)}_{31}(\tau)$. 
The function $g^{(2)}_{ii}(\tau)$ can be tailored to show sub-poissonian statistics
by making the coupling field  
weak ( $\Omega_1<\bar \Gamma_2)$ . The auto-correlation function $g_{33}^{(2)}(\tau)$
shows sub-poissonian statistics even for $\Omega_3> \bar \Gamma_4$ since the population of the excited levels is smaller.
Fig4 shows the correlation functions and the ratio $R$ for various parameter values. 
Here, the dimensionless variable $t=\gamma \tau$ where $\gamma= 2 \pi MHz$.
The function $g^{(2)}_{31}(\tau)$ increases with $\Omega_{rf}$ (Fig4 (c)),
while both the auto-correlation functions decrease with $\Omega_{rf}$
 (Fig4 (a) and (b)). 
On the other hand, the time delay for the occurrence of peak value
of $g_{ij}^{(2)}(\tau)$ is smaller for larger values of $\Omega_{rf}$. This means that, in principle,
the occurrence of $R^{max}$ can be delayed/advanced by decreasing/increasing
the coupling field strengths - implying a trade-off between the
value of $R^{max}$ and the time delay for the occurrence of the same.
 Note that  $R^{max}=10^3-10^6$ as shown in Fig4 (d)  and increases with
$\Omega_{rf}$ . It is noteworthy that the oscillatory behaviour of $R$ for
 large $\Omega_{rf}$ provides different time delays at which the violation
is maximum. 

Lastly, the merit in this scheme is that the 
the absorption profile being very narrow (of the order of the natural line-width) \cite{sns}, the excitation/deexcitation
of the g.s. and the upper excited state occurs within a very narrow frequency
band. 
This is because of the three photon resonance which induces atomic coherence. 
It has been predicted earlier \cite {sandhya} 
 that when a three-level
cascade system showing EIT is coupled to a third driving field, the ground state (g.s.)
shows a very narrow absorption within the EIT window at three photon resonance; 
the Doppler integrated absorption does not suffer much change in this geometry.
 This type of  narrow absorption features and three-peak absorption,
 in four-level systems, have been subsequently reported experimentally \cite{suter}. 
An insight into this behavior is revealed by the study of the single atom dynamics \cite{sns07}: when the middle transition
coupling is strong ($\Omega_{rf}$ in this case) the four level atom behaves 
like a two level atom spanned by levels $|1>$ and $|4>$ hence inducing
a strong atomic coherence between these two levels. A steady state analysis
\cite{sns} further
reveals that in an ensemble
this feature manifests as 
a dominant contribution of the third order nonlinear susceptibility (due to the
strong atomic coherence)
as compared to the linear susceptibility.
This third order nonlinear susceptibility is responsible for the efficient
transfer of population to the uppermost level by inducing a narrow
absorption in the ground state. The role of $\Omega_{rf}$ is not only to 
transfer the population to level $|4>$ but the strong atomic coherence
it introduces is responsible for the strong correlation between the
photons emitted by the extreme transitions.

The absorption features for the model under consideration is very similar, 
and since the absorption occurs within an EIT window, there is an absorption free zone \cite{sns} for
detunings larger than the line width of the central peak,
and detunings less than the FWHM of the EIT window. i.e. for $ 15 MHz >| \Delta_1| > 6 MHz $. 
This ensures cleaner signals and  a reduction in the noise due to the absorption and emission into other modes close to the resonant frequency which is a major advantage in this system. 
I would like to mention here that if the coupling between the 
levels $|2>-|3>$ were in the visible region instead of radio-frequency region
then the absorption of the $|3>-|4>$ transition diminishes 
due to Doppler averaging \cite{sandhya}. Even though this does not affect
the antibunching nature of $g^{(2)}_{31}(\tau)$, the strength of the correlation function
would become weaker. Hence the signature for the violation
of the Cauchy-Schwarz inequality may also become weaker.

To summarize, I propose a model for generating non-classical pairs, wherein
Doppler effects are negligible at room temperatures, and also the absorption/emission 
proceeds in a narrow frequency band .
The violation of the Cauchy-Schwarz inequality is orders of magnitude larger
than the experimentally reported data, so far, and therefore could provide
very clear signals in the near future. Another interesting feature is the 
antibunching nature of the cross-correlation between the emission of photons from the
extreme transition which occurs because of the cancellation of emission
due to different two-photon pathways. Thus the photon pair generated in this
scheme 
shows the violation of two of the classical inequalities. 
\section*{Appendix}
The solutions for the density matrix elements in the Laplace space
for various initial conditions
and both the strong coupling of $\Omega_{rf}$ and weak $ \Omega_{rf}$ are
listed below.
\subsubsection*{Strong $\Omega_{rf}$}
For the initial condition $\rho(0)=|3><3|$, the relevant $\rho_{ij}(\tau)$ in the Laplace
space is given by
\begin{eqnarray}
{\cal L}(\rho_{22}(\tau))=& \bar \psi_7(s)\nonumber \\
&= \frac{1}{d_3} [ (s+\bar \Gamma_2+\bar \Gamma_3 +2 \Omega_2^2)  (1+
\frac{2 \Omega_3^2}{d_4 (s+\bar \Gamma_4)}
(1-s
\nonumber \\&
-\bar \Gamma_4)(
s+\bar \Gamma_2+\bar \Gamma_4 )) +\frac{2}{d_4}\Omega_2^2 \Omega_3^2(1-s-\bar \Gamma_3)]\nonumber 
\end{eqnarray}
where $d_4=\Pi_{i=1}^2(s+\bar \Gamma_4+\alpha_i), d_3=\Pi_{i=3}^5(s+\alpha_i),
d_2=\Pi_{i=1}^2(s+\alpha_i)$
\begin{eqnarray}
{\cal L}(\rho_{33}(\tau))=& \bar \psi_8(s)\nonumber \\
&= \frac{1}{d_3}[ (s^2+\bar \Gamma_2^2+\bar \Gamma_2\bar \Gamma_3+
s(\bar \Gamma_3+2\bar \Gamma_2)
\nonumber \\ &
+ 2 \Omega_2^2)
 (1+ \frac{2\Omega_3^2}{d_4 (s+\bar \Gamma_4)}
(s+\bar \Gamma_2+\bar \Gamma_4)
\nonumber \\ &
(1-s-\bar \Gamma_4))
+ \frac{1}{d_4}2\Omega_2^2\Omega_3^2(s+\bar \Gamma_2)]\nonumber 
\end{eqnarray}

For the initial condition $\rho(0)=|2><2|$ we have 
\begin{eqnarray}
{\cal L}(\rho_{22}(\tau))=& \bar \psi_7(s)\nonumber \\
&= \frac{1}{d_3}[ (s^2+\bar \Gamma_3^2+s(\bar \Gamma_2+2 \bar \Gamma_3)+
\bar \Gamma_2\bar \Gamma_3+
\nonumber \\ &
2\Omega_2^2)
+\frac{1}{s d_2 } 
(2\Omega_1^2\Omega_2^2(s-1) 
 -2\Omega_1^2(s+\bar \Gamma_3)
\nonumber \\ &
(s^2+\bar \Gamma_3^2+s(\bar \Gamma_2+2\bar \Gamma_3)+
2\Omega_1^2+\bar \Gamma_3 \Omega_2^2))]\nonumber 
\end{eqnarray}

For the initial condition $\rho(0)=|1><1|$ we get
\begin{eqnarray}
{\cal L}(\rho_{22}(\tau))=&\bar \psi_7(s)\nonumber \\
&= \frac{2\Omega_1^2}{s d_2 d_3}[ \Omega_2^2+(s+\bar \Gamma_3)((s+\bar \Gamma_3+\bar \Gamma_2)
\nonumber \\ &(s+\bar \Gamma_3)+\Omega_2^2)]\nonumber 
\end{eqnarray}
\subsubsection*{ Weak $\Omega_{rf}$}
For the initial condition $\rho(0)=|3><3|$ we have
\begin{eqnarray}
{\cal L}(\rho_{22}(\tau))= &\bar \psi_7(s)\nonumber \\
&= \frac{1}{d_{2p}}[\frac{2 \Omega_1^2}{s}+ 2 \Omega_1 \Omega_2 {\cal C}_1-
2(s+\bar \Gamma_2)\Omega_2 {\cal C}_2+
\nonumber \\ &
\frac{4 \Omega_1^2}{d_{3p}} (-\Omega_3^2+
 \Omega_2\Omega_3
( (s+\bar \Gamma_3) {\cal C}_3- 
2 \Omega_3{\cal C}_2))\nonumber \\
&+
\frac{1}{(s+\bar \Gamma_2) d_{3p}} (\bar \Gamma_2^2+s(s-2\Omega_1^2)+
2\bar \Gamma_2(s-\Omega_1^2))
\nonumber \\ &
( (1+2 \Omega_2 {\cal C}_2)
(s^2+\bar \Gamma_4^2+\bar \Gamma_3(s+\bar \Gamma_4)+
\nonumber \\&
2 \bar \Gamma_4 s+2 \Omega_3^2)+
2{\cal C}_3 \Omega_2\Omega_3(s+\bar \Gamma_4-1))]\nonumber
\end{eqnarray} 
\begin{eqnarray}
{\cal L}(\rho_{33}(\tau))=&\bar \psi_8(s)\nonumber \\&
=\frac{1}{d_{3p}}[(1+2 \Omega_2 {\cal C}_2)(s^2+\bar \Gamma_4^2+\bar \Gamma_3
(s+\bar \Gamma_4)+
\nonumber \\ &
2\Omega_3^2+
2 s \bar \Gamma_4)+ 
2 \Omega_2\Omega_3 {\cal C}_3(s+2\bar \Gamma_4+1)]\nonumber
\end{eqnarray}
\begin{eqnarray}
{\cal L}(\rho_{44}(\tau))=& \psi_9(s)\nonumber \\ &=
\frac{2\Omega_3}{d_{3p}}[\Omega_3(1+2 \Omega_2 {\cal C}_2)-\Omega_2 {\cal C}_3
(s+
\bar \Gamma_4)] \nonumber
\end{eqnarray}

For the initial condition $\rho(0)=|2><2|$ we have
\begin{eqnarray}
{\cal L}(\rho_{22}(\tau))= &\bar \psi_7(s)\nonumber \\
&= \frac{1}{d_{2p}}[\frac{2 \Omega_1^2}{s}+ 2 \Omega_1 \Omega_2 {\cal C}_1-
(s+\bar \Gamma_2)
(1-
\nonumber \\ &
2\Omega_2 {\cal C}_2)
+ \frac{4 \Omega_1^2\Omega_2^2\Omega_3^2}{d_{3p}} (
( (s+\bar \Gamma_3) {\cal C}_3-
2 \Omega_3{\cal C}_2))
\nonumber \\ & 
+\frac{1}{(s+\bar \Gamma_2) d_{3p}} (\bar \Gamma_2^2+
s(s-2\Omega_1^2)+
2\bar \Gamma_2(s-\Omega_1^2)
\nonumber \\ &
( 2 \Omega_2 {\cal C}_2
(s^2+\bar \Gamma_4^2+
\bar \Gamma_3(s+\bar \Gamma_4)+
2 \bar \Gamma_4 s
\nonumber \\ &
+2 \Omega_3^2)
+2{\cal C}_3 \Omega_2\Omega_3(s+\bar \Gamma_4-1)))]\nonumber
\end{eqnarray} 

For the initial condition $\rho(0)=|1><1|$
\begin{eqnarray}
{\cal L}(\rho_{22}(\tau))=&\psi_7(s)= 2\Omega_1^2/(s d_{2p})\nonumber
\end{eqnarray}
where 
\begin{eqnarray}
{\cal C}_1=&-\Omega_1\Omega_2[(s+\bar \Gamma_4)(s+\bar \Gamma_4+
\nonumber \\ &
\bar \Gamma_2)+\Omega_1^2-\Omega_3^2]/d_{4p}\nonumber \\
{\cal C}_2=&-\Omega_2[ (s+\bar \Gamma_4)(s+\bar \Gamma_3)(s+\bar \Gamma_4+
\bar \Gamma_2)+\nonumber \\ &
\Omega_3^2(s+\bar \Gamma_2+ \bar \Gamma_4)+
(s+\bar \Gamma_3)\Omega_1^2)]/d_{4p}\nonumber \\ 
{\cal C}_3=& \Omega_2\Omega_3[-(s+\bar \Gamma_3)(s+\bar \Gamma_4)+
\nonumber \\ &
\Omega_1^2-\Omega_3^2]/d_{4p}\nonumber
\end{eqnarray}

Here $d_{4p}=\Pi_{i=6}^9 (s+\bar \alpha_i)$, $ d_{3p}= \Pi_{i=3}^5 (s+\bar \alpha_i)$ and $d_{2p}=(s+\bar \Gamma_2)^2+4\Omega_1^2$.

\section*{Acknowledgments}
I wish to thank the Department of Science and Technology, INDIA,
 for Financial
support under the WOS-A scheme.
 

\end{document}